\documentclass{article}

\usepackage{arxiv}

\usepackage[utf8]{inputenc} 
\usepackage[T1]{fontenc}    
\usepackage{url}            
\usepackage{booktabs}       
\usepackage{amsfonts}       
\usepackage{nicefrac}       
\usepackage{microtype}      
\usepackage{lipsum}
\usepackage{graphicx}
\graphicspath{ {./images/} }

\usepackage{amsmath,amsfonts,amssymb}
\usepackage{graphicx}
\usepackage{soul}
\DeclareUnicodeCharacter{2061}{}
\usepackage[utf8]{inputenc}
\usepackage{newunicodechar}
\usepackage[colorlinks=true, allcolors=blue]{hyperref}
 \newcommand{\ket}[1]{\left| #1 \right>}

\title{Randomized ancillary qubit overcomes detector-control and intercept-resend hacking\\ of quantum key distribution}

\author{
 Salem~F.~Hegazy \\
  National Institute of Laser Enhanced Sciences, \\ 
  Cairo University, \\
  Giza 12613, Egypt\\ \texttt{salem@niles.cu.edu.eg}\\
   \And
 Salah S. A. Obayya \\ 
Center for Photonics and Smart Materials,\\
Zewail City of Science and Technology,\\ Giza 12578, Egypt \\
 \texttt{sobayya@zewailcity.edu.eg}\\
  \And
 Bahaa E. A. Saleh\\ 
 CREOL, The College of Optics \& Photonics,\\
 University of Central Florida,\\ Orlando, FL, 32816, USA \\
  \texttt{besaleh@creol.ucf.edu}\\
}

\begin{document}
\maketitle
\begin{abstract}
Practical implementations of quantum key distribution (QKD) have been shown to be subject to various detector side-channel attacks that compromise the promised unconditional security. Most notable is a general class of attacks adopting the use of faked-state photons as in the detector-control and, more broadly, the intercept-resend attacks. In this paper, we present a simple scheme to overcome such class of attacks: A legitimate user, Bob, uses a polarization randomizer at his gateway to distort an ancillary polarization of a phase-encoded photon in a bidirectional QKD configuration. Passing through the randomizer once on the way to his partner, Alice, and again in the opposite direction, the polarization qubit of the genuine photon is immune to randomization. However, the polarization state of a photon from an intruder, Eve, to Bob is randomized and hence directed to a detector in a different path, whereupon it triggers an alert. We demonstrate theoretically and experimentally that, using commercial off-the-shelf detectors, it can be made impossible for Eve to avoid triggering the alert, no matter what faked-state of light she uses.
\end{abstract}


\section{Introduction}
%
%
%
%
The unconditional security offered by quantum key distribution (QKD) relies on laws of quantum physics \cite{BB84, Ekert91}, which dictate that any attempt by an adversary to know about the secret key, would inevitably introduce disturbance that alerts the legitimate parties \cite{Bennett94, Shor-Preskill}. This ultimate information-theoretic security has been proved for idealized devices \cite{Shor-Preskill,Mayers01,Renner08} and also under semi-realistic conditions \cite{Gottesman04_GLLP,Inamori07,Tomamichel12}. In practice, however, real-life components of QKD systems may deviate from these idealized theoretical models, or encounter new scenarios, offering effective vulnerabilities to the adversary. 

For instance, the imperfect preparation of the single-photon state may lead to leaking information about the key. This gap between theory and real-life practice allows for a plethora of source-side attacks ranging from the photon-number-splitting (PNS) attack \cite{Brassard00, Lutkenhaus00}, the phase-remapping attack \cite{Fung07,Xu10}, the wavelength-selected photon-number-splitting attack \cite{Jiang12}, and the pattern-effect attack \cite{Yoshino18}, to the nonrandom-phase attacks based on unambiguous-state-discrimination \cite{Tang13}, and laser seed control \cite{Sun15, Pang20, Huang19}.

Compared to the source-side attacks, imperfections on the detection side are known to show much higher vulnerability to quantum hacking \cite{Xu20}. For example, detector imperfections such as breakdown fluorescence \cite{Newman55}, finite ($\sim\mu$s) dead time \cite{Henning_Weier11}, nonzero dark counts, less-than-unity efficiency, and nonfixed efficiency within the gate time \cite{Zhao08}, all of which can be exploited by Eve to compromise QKD security. This leads in practice to a significant number of potential attacks such as  detector fluorescence \cite{Kurtsiefer01}, faked-state \cite{Makarov06, Makarov05}, time-shift \cite{Qi07, Zhao08}, time-side-channel \cite{Lamas07}, channel calibration \cite{Jain11}, laser damage \cite{Bugge14, Makarov16}, spatial mismatch \cite{Chaiwongkhot19, Sajeed15}, detector saturation \cite{Qin16}, and polarization shift \cite{Wei19} attacks. More interestingly, the single-photon detectors (SPDs) of the receiver (Bob), normally operating in the Geiger mode \cite{Hadfield09}, can be turned by Eve into linear mode, which allows for various blinding and remote-control attacks \cite{Makarov09, Lydersen_Nat10, Gerhardt11a, Gerhardt11b, Lydersen_NJP_11, Wiechers11, Qian18, Wu20}. Among the detection-side attacks, the latter is widely known to be the most powerful \cite{Xu20}, with successful demonstrations on various types of SPDs, including passively and actively quenched avalanche photodetectors (APDs) \cite{Makarov09, Sauge11}, gated/non-gated APDs \cite{Huang16, Lydersen_Nat10}, and superconducting nanowire single-photon detectors (SNSPDs) \cite{Lydersen-PRA-11}.

Since the inception of quantum encryption \cite{BB84}, the intercept-resend strategies have been developed through many quantum hacking paradigms. Its original version based on resending single photons was easily neutralized by QKD \cite{Bennett94}. Employing detector imperfections, more crafty intercept-resend versions have evolved via resending faked multiphoton states either solitarily (e.g., the after-gate attack \cite{Wiechers11}, the faint-after-gate attack \cite{Lydersen-PRA-11}, and the detector-control attack under specific laser damage \cite{Bugge14}) or teamed with a blinding light (e.g., continuous-wave blinding attack \cite{Lydersen_Nat10, Gerhardt11a}, sinkhole blinding attack \cite{Lydersen-Opt10}, thermal blinding attack \cite{Lydersen-Opt10, Sauge11}, and pulsed illumination attack \cite{Wu20}).

Currently, there exist two main approaches against the intercept-resend and detector-control hacking strategies. The first is based on monitoring some detector measures, such as its photocurrent, for anomalously excessive values \cite{Yuan10, Yuan11, Silva12}. This includes also observing the detector’s count rates versus random variations of either the detection efficiency \cite{Patent, Lim15}, or the attenuation in front of the detector \cite{Qian19}. These security patches could defeat the original attacks they were designed for, but unfortunately they fail against subsequent \textit{ad-hoc} modified attacks \cite{Huang16,Wu-comment20}.

The second is the measurement-device-independent QKD (MDI-QKD) approach \cite{Lo12}, which enables elimination of all detector side-channels \cite{Braunstein12}, offering security regardless of the nature of the detection apparatus. However, MDI QKD builds on performing a remote Bell-state measurement, which requires high-visibility two-photon interference between independent photons from Alice’s and Bob’s laser sources, a practically challenging procedure.
 
In this paper, we present a scheme to protect practical QKD systems against various attacks based on faked-state light, including the detector-control attacks and more generally the class of intercept–resend attacks. The scheme uses phase encoding and a two-way configuration, similar to the plug-and-play configuration \cite{Stucki02,Bethune02,Park18}, which uses polarization-assisted routing through Bob’s transceiver, and a Faraday mirror at Alice’s site.  In our scheme, however, the polarization qubit serves a different function.  A photon generated at Bob’s transceiver is transmitted through a polarization randomizer, which assigns it a random state of polarization, and upon reflection from the Faraday mirror it passes once more through the same randomizer, in a state orthogonal to its original state, and is directed to a specific path, whereupon the photon is detected in accordance with the phase-encoded BB84 protocol.  Light pulses generated by an intruder must pass through the randomizer at the gateway to Bob’s transceiver, and since they pass only once, they acquire a random state and end up in a different path, whereupon their detection triggers an alert.   The randomizer is fixed during the course of the photon roundtrip and is refreshed after every cycle of photon transmission and detection. Thus, the polarization qubit serves as a carrier of a \textit{password} that allows genuine photons to be directed to the secured detectors, while an intruder’s fake photons are randomized and possibly end up at the alert detectors.

We further consider the case that Eve launches a generalized detector-control attack. To render her attack unnoticeable, she tailors the parameters of triggering pulses and blinding light in order to meet two requirements: (i) to avoid triggering alert detectors, and (ii) to be able to sometimes trigger the secured detectors in the right way.
These two requirements lead us to a necessary and sufficient condition that Bob's secured and alert detectors have to satisfy. 
We note that commercially available detectors can violate this necessary and sufficient condition and thereby guarantee that these two requirements are impossible to meet simultaneously. We experimentally demonstrate how various faked states by Eve fail to simultaneously meet these two requirements of unnoticeable attack. {Security analysis of the system shows that for various types of attacks Eve cannot diminish the alert rate, even if she has complete control over Bob's secured detectors.}


 \begin{figure*} [t!]
   \begin{center}
   \includegraphics[width=\linewidth]{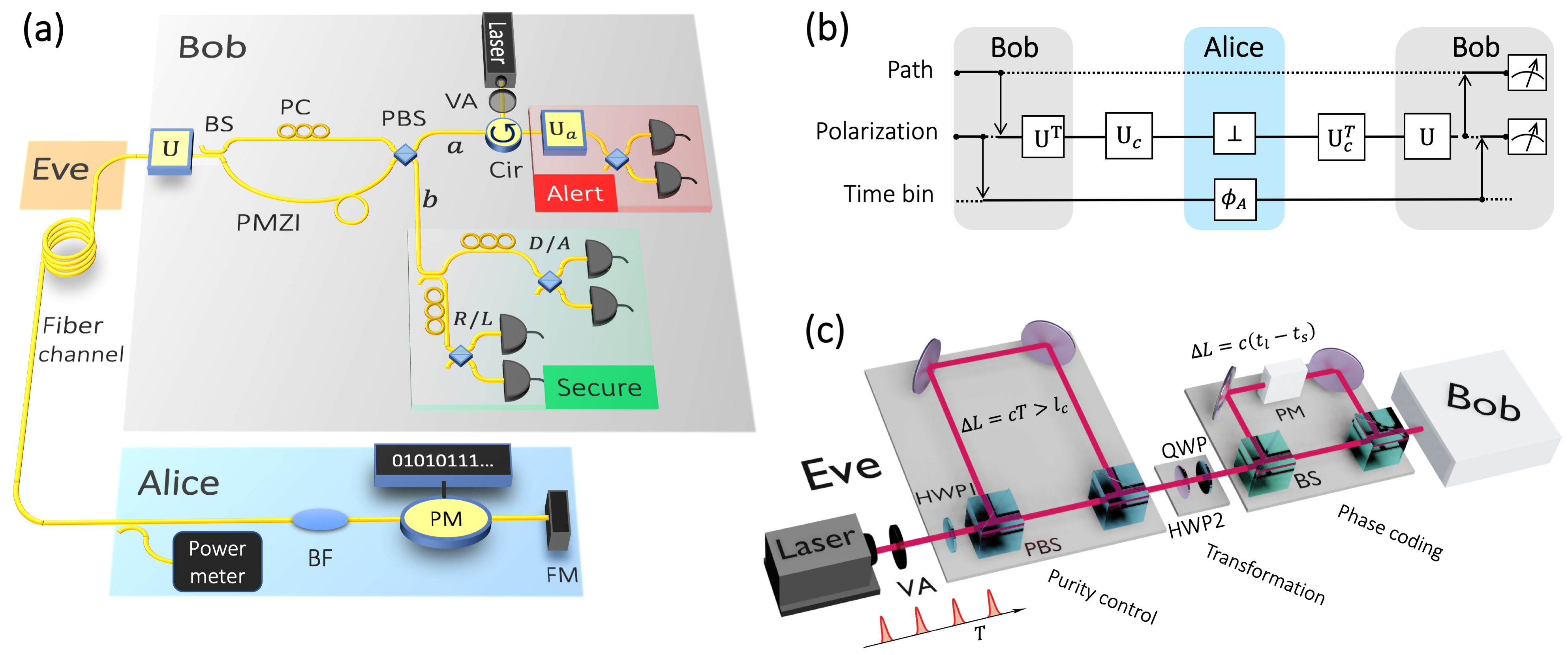}
   \end{center}
   \caption[example] 
   { \label{fig_setup} 
\textbf{(a)} Optical layout of the QKD system. Bob creates single photons with time-bin (key) and polarization (ancillary) qubits. The polarization qubit is randomized by an operator $\mathrm{U}$, only known to Bob. Alice’s phase modulator (PM) encodes the time-bin state by a phase $\phi_A \in \{0,\pi\}$ or $\{\pi/2, 3\pi/2\}$. A Faraday mirror (FM) compensates Bob’s back-tracing photon for all encountered polarization variations, including the randomization $\mathrm{U}$. The polarization-based Mach-Zehnder interferometer (PMZI) swaps the time-bin/polarization qubits for polarization/path qubits. Therefore, the key qubit is measured in path $b$ in either diagonal-antidiagonal ($D/A$) or right-left ($R/L$) circular polarization bases. The polarization randomizer $\mathrm{U}$ {– which may be implemented by means of high-speed electro-optic polarization controller –} is active against Eve’s fake photons and may direct them, without Eve’s notice, to the alert detectors in path $a$. A click of the alert detectors in path $a$ is a sign for Eve’s intrusion. The polarization switch $\mathrm{U}_a$ alternates between measurements bases: $D/A$ and $R/L$. BS: beam splitter; PBS: polarization beam splitter; PC: polarization controller; Cir: optical circulator; VA: variable attenuator; BF: narrow band-pass filter. \textbf{(b)} The timeline for the operations on qubits of the three photonic degrees of freedom, path, time bin, and polarization, during the course of a roundtrip from Bob to Alice and back along a channel $\mathrm{U}_c$. {The operator $\mathrm{U}$ describes the polarization transformation, when light enters Bob’s system. In the opposite direction, it encounters a transformation $\mathrm{U}^T$.}  \textbf{(c) }Optical setup demonstrating Eve’s system. The half-wave plate HWP1 and the following polarization-based two-path system control the purity of the polarization state via mixing orthogonal polarization components of two subsequent laser pulses. {The subsequent half- and quarter-wave plates, HWP2 and QWP, alter the polarization state unitarily. }The two-path system in the last stage performs the time-bin phase encoding.}
 \end{figure*} 

\section{QKD scheme}
As shown in Fig. \ref{fig_setup}(a), Bob employs a single photon with two encoded qubits: a time-bin qubit communicating the \textit{key}, and an ancillary polarization qubit serving as a \textit{security pass} \cite{ourpatent}.  As in typical interferometric QKD systems, the photon undergoes a roundtrip from Bob to Alice, where the time-bin qubit is modulated, and sent back to Bob whereupon it is directed to two sets of detectors depending on its state of polarization. {Entry into Bob’s receiver is secured by a polarization randomizer applying a random transformation $\mathrm{U}$ (based on Haar measure) that changes every photon-roundtrip duration.} Alice uses a Faraday mirror (FM) that switches the polarization qubit into an orthogonal state so that as the photon crosses the polarization randomizer in the opposite direction, the randomization is cleared. Since its state is only known to Bob, the randomizer is a secure polarization-based gateway that directs the photon to specific detectors in the receiver.

The process begins as shown in Fig. \ref{fig_setup}(a) with Bob sending single-photon pulses along path $a$ in a polarization-path state: 
\begin{equation}
\ket{\psi_1}={1}/{\sqrt{2}}(\ket{H}+\ket{V})\ket{a}. 
\end{equation}
This is subsequently swapped for a time-polarization state 
\begin{equation}
\ket{\psi_2}={1}/{\sqrt{2}}(\ket{t_l}+\ket{t_s})\ket{H}
\end{equation}by use of an unbalanced polarization-based Mach-Zehnder interferometer (PMZI) with a polarization controller (PC) placed in its short arm, converting the V (H) polarization into H (V) polarization. 

On Alice’s side, the leading time bin $\ket{t_s}$ is encoded with a phase shift $\phi_A$ of $0$ or $\pi$, and $\pi/2$ or $3\pi/2$. Upon reflection from the FM, the photon polarization is flipped to its orthogonal state. This compensates for the undesired polarization changes accompanying the phase modulation \cite{Muller97}, and also for the birefringence-based polarization fluctuations along the optical fiber \cite{Gisin02, Souto19}. Upon re-entry into Bob’s transceiver, since $\mathrm{U}$ is fixed during the photon roundtrip, its effect is also cancelled out by transmission in the opposite direction. The state is now: 
\begin{equation}
\ket{\psi_3}=\tfrac{1}{\sqrt{2}}(\ket{t_l}+e^{i\phi_A}\ket{t_s})\ket{V}.
\end{equation}
Bob’s receiver is gated to select roundtrip passage via the short-long and the long-short arms of the PMZI arms. It is also configured such that with single-photon interference in the PMZI, the time-polarization state $\ket{\psi_3}$ is swapped back to a polarization-path state 
\begin{equation}
\ket{\psi_4}={1}/{\sqrt{2}}(\ket{H}+e^{i\phi_A}\ket{V})\ket{b}.
\end{equation}
The photon is therefore directed to path $b$, which we call the \textit{secure} path. As will be shown later, detection of a photon in path $a$ is an indication that the system has been tampered with, and path $a$ is therefore called the \textit{alert} path.

After swapping the key qubit back to polarization, the BB84 measurement is performed passively in one of the conjugate bases: diagonal-antidiagonal ($D/A$) or right-left ($R/L$) circular polarization. The system’s action on the different degrees of freedom (path, time, and polarization) of the photon during its roundtrip course is illustrated in Fig. \ref{fig_setup}(b).

{In yet another measure of added security, Bob randomly directs the received photon – in a managed way – to path $a$ instead of path $b$ for measurement. This is accomplished by appropriate control of the polarization randomizer. This random-switching tactic unveils types of attacks that can bias triggering actions to path $b$ such as pulsed-blinding \cite{Wu20, Wu-comment20, Henning_Weier11} and wavelength-dependent attacks \cite{Li2011}.}

Alice's phase coding and Bob's gated detection require precise time synchronization between the two sides which is done via a wavelength-multiplexed classical channel carrying bright pulses. A portion of the power received by Alice is monitored to detect Trojan horse attacks \cite{Gisin02}.

Here, an ideal single-photon source is assumed for convenience. To defend against the PNS attack, Bob applies a decoy-state technique \cite{Hwang03,Lo05, Wang05}; verifying that his produced decoy pulses encounter the same single-photon loss. 

{
\section{Randomized routing of faked-state light }
}
 Eve's goal is to signal the detectors in the secure path $b$ without registering a click on the detectors of the alert path $a$. In a typical intercept–resend strategy, Eve would measure Alice's encoded state and then send faked-state light in a phase modulated state $(\ket{t_l}+e^{i\phi_E}\ket{t_s})/{\sqrt{2}}$, mimicking the measured key qubit, together with a polarization qubit in a state $\rho_p$. Upon transmission through the PMZI, and within the detection window (centered at: $t_s+t_l$), the state of Eve’s photon(s) becomes
\begin{equation}
\begin{split}
\tfrac{1}{2}  |H\rangle \langle H|&(\ket{b}X + e^{i\phi_E} \ket{a}) \mathrm{U} \rho_p \mathrm{U}^+ \\
& \times (X\langle b|+ e^{-i\phi_E} \langle a|)|H\rangle \langle H|\\
+ \tfrac{1}{2}  |V\rangle \langle V|&(\ket{a}X + e^{i\phi_E} \ket{b}) \mathrm{U} \rho_p \mathrm{U}^+ \\
& \times (X\langle a|+ e^{-i\phi_E} \langle b|)|V\rangle \langle V|.
\end{split}
\end{equation}
The NOT operator $X$ is due to action of the PC in the PMZI. To obtain the which-path statistics, we trace over polarization and obtain the reduced density operator of the path states 
               \begin{equation}
               \begin{split}
                             & p_a |a\rangle \langle a| + \cos \phi_E  \langle H|\mathrm{U} \rho_p \mathrm{U}^+ |V\rangle  |a\rangle \langle b| \\
                             + &\cos \phi_E  \langle V|\mathrm{U} \rho_p \mathrm{U}^+ |H\rangle |b\rangle \langle a|+ p_b |b\rangle \langle b| .
               \end{split}
               \label{Eqwhichpath}
               \end{equation}
The probabilities that Eve’s photon(s) ends up in the alert path $a$ is $p_a=\langle H|\mathrm{U} \rho_p \mathrm{U}^+ |H\rangle$, while that of reaching path $b$ is  $p_b=1-p_a=\langle V|\mathrm{U} \rho_p \mathrm{U}^+ |V\rangle$. If Eve were to know the operator $\mathrm{U}$, she would be able to make $p_a= 0$ by use of a pure state $\rho_p=\mathrm{U}^+ |V\rangle \langle V|\mathrm{U}$.  
Not knowing $\mathrm{U}$, if she runs the conventional intercept-resend attack \cite{BB84, Bennett92} by measuring the Alice-encoded photon and re-sending a new photon prepared in accordance with the measurement outcome to Bob, then the average probability that it passes to path $a$ is 25\% (obtained by averaging over the continuum of random realizations of $\mathrm{U}$ based on Haar measure, assuming ideal single-photon sources, measurements, and detection). This alert rate is on top of the normal 25\% quantum bit error rate (QBER) of the BB84 key qubit.
{
\section{Necessary criteria for Bob's detectors }
\subsection{Criteria formulation}
}
A more stealth intercept–resend strategy that we now investigate in more details is Eve’s use of blinding light together with triggering multi-photon pulses  \cite{Lydersen_Nat10, Gerhardt11a, Lydersen-Opt10, Sauge11, Wu20}. Upon blinding, the SPD in the linear mode never clicks when the triggering pulse energy is below a threshold $E_{never}$, and always clicks when the energy is greater than a threshold $E_{always}$ \cite{Huang16, Wu20}. When the energy falls between these two levels, the detector clicks with a probability between 0 and 1.  

For hacking the BB84 QKD system, it is required that $E_{always}<2E_{never}$ so that if the trigger pulse has energy $E_T  \in [E_{always}, 2 E_{never})$, the detector will always click in the compatible basis, but will never click in the conjugate basis. Bob’s detectors can then be fully controlled without elevating the QBER \cite{Lydersen_Nat10}.

The reason for the potential of this hacking strategy is shown by noting that upon blinding, the alert SPDs in path $a$ will receive double the blinding power –on average– relative to the SPDs in path $b$ [see Fig. \ref{fig_setup}(a)]. Because $E_{always} (I)$ and $E_{never} (I)$ are monotonic increasing functions of the blinding power $I$ \cite{Huang16}, higher blinding power for the alert SPDs generally elevates their operation thresholds. As a result, one might think that the alert SPDs would be more insensitive to the triggering pulses, which could be exploited to produce an unnoticeable intrusion. 

To investigate this attack further, let us consider that Eve uses triggering pulses of energy $E_T$ carrying her measured key (time-bin) qubit together with an ancillary (polarization) state $\rho_T$. This is accompanied by blinding light of power $I_B$ and polarization state $\rho_B$. Eve would like to optimize the attack parameters — $E_T,\rho_T,I_B$, and $\rho_B$— aiming to perform selective triggering of detectors in path $b$ without registering a click in the alert SPDs in path $a$. In the following analysis, we show that such goal can be made impossible if Bob’s SPDs are appropriately selected. 

Eve’s photons of the trigger pulse will be split into paths $a$ and $b$ with the probabilities $p_a$ and $p_b$, and then split again equally between the two polarization paths of $b$. If the total energy of Eve’s time-bin pulses is $E_T$, then within the gated time window there will be a portion $\tfrac{1}{2} p_a E_T$ in path $a$ (this is also the maximum energy received by any detector $D_{ai}, i \in \{1,2\}$), and portions  $\tfrac{1}{4} p_b E_T$ in each arm of path $b$ (the maximum energy received by any detector $D_{bj}, j \in \{1,2,3,4\}$).  

To develop a successful detector control, Eve’s triggering pulse and blinding light have to satisfy concurrently the following two conditions for all possible realizations of $\mathrm{U}$:

 \textbf{(A)}: The maximum trigger pulse energy that may strike an alert detector $D_{ai}$ is less than the minimum $E_{never}^{ai}$, i.e.,
 \begin{equation}
 \tfrac{1}{2} p_{\max} E_T < E_{never}^{ai}(\min \{I_a\}),
 \label{Cond:A}
 \end{equation}
where $\min \{I_a\}$ is the minimum blinding power received by a detector $D_{ai}$ and $p_\mathrm{max}$ is the maximum value of $p_a$ obtained over any state $\mathrm{U}\rho_T\mathrm{U}^+$ (see Appendix \ref{APP:C:Overlap}), which is given by  
\begin{equation}
p_\mathrm{max} = \tfrac{1}{2}(1+\sqrt{2 \mathcal{P}_T-1}),
\label{eq:pmax}
\end{equation}
with $\mathcal{P}_T$  being the purity of the polarization state $\rho_T$.  

 \textbf{(B)}: The maximum pulse energy that may strike a detector $D_{bj}$ must be at least greater than the minimum $E_{never}^{bj}$, i.e.,
  \begin{equation}
 \tfrac{1}{4} p_{\max} E_T > E_{never}^{bj}(\min \{I_b\}),
 \label{Cond:B}
 \end{equation}
where $\min \{I_b\}$ is the minimum blinding power received by a detector $D_{bj}$, and $p_\mathrm{max}$ is the maximum of $p_b$ taken over any state  $\mathrm{U}\rho_T\mathrm{U}^+$ [same as in (\ref{eq:pmax})].

Condition (A) guarantees that even if the maximum triggering-pulse energy passes to a detector $D_{ai}$, which is blinded with the minimum light power, this should not lead to a click. {Condition (B) offers a necessary condition for detectors $D_{bj}$ to trigger. }

As shown by (\ref{Cond:A}) and (\ref{Cond:B}), for Eve who does not know about the transformation $\mathrm{U}$, the maximum pulse energy (over all possible settings of $\mathrm{U}$) that may impinge on a detector $D_{bj}$ is half that for an alert detector $D_{ai}$. Consequently, conditions (A) and (B) cannot be satisfied unless the detectors $D_{ai}$ and $D_{bj}$ strictly comply with the necessary and sufficient condition:
\begin{equation}
\frac{E_{never}^{bj}(\min \{I_b\})}{E_{never}^{ai}(\min \{I_a\})} < \frac{1}{2}, ~~\forall ~i,j
\label{eq:cond1/2}
\end{equation}
Because Eve does not know the current $\mathrm{U}$, she does not have the ability to reliably control the ratio of the maximum pulse energies delivered to the detectors $D_{ai}$ and $D_{bj}$. Thus, Bob’s setup restricts this ratio in operation to ½ as in (\ref{Cond:A}) and (\ref{Cond:B}) due to the balanced beamsplitting in path $b$.

Aiming to avoid the alert SPDs in path $a$, Eve will gain no benefit by assigning a specific time-bin state for the blinding light. We therefore assume, without loss of generality, that the blinding light is in a mixed time-bin state. For an input blinding light of power $I_B$ and a state of polarization $\rho_B$, the power received by the SPDs $D_{ai}$ and $D_{bj}$ are, respectively, $I_a=\tfrac{1}{2} r_a I_B$ and $I_b= \frac{1}{4} r_b I_B$, where $r_a=\langle H|\mathrm{U} \rho_B \mathrm{U}^+ |H\rangle$ and $r_b=\langle V|\mathrm{U}\rho_B \mathrm{U}^+ |V\rangle$.
The probabilities $r_a$ and $r_b$ are bounded over all settings of $\mathrm{U}$ by the same minimum value: $\tfrac{1}{2}(1-\sqrt{2 \mathcal{P}_B-1})$ (see Appendix \ref{APP:C:Overlap}), where $\mathcal{P}_B$ is the purity of the state $\rho_B$. It follows that: 
\begin{equation*}
\frac{\min \{I_b\} }{\min \{I_a\}} =\frac{1}{2}.
\end{equation*}
Note that the variations in $\mathrm{U}$ for each roundtrip alters the value of the blinding power illuminating the SPDs. Here, we assumed that the threshold $E_{never}$ depends on the instantaneous blinding power. However due to the electronics of the SPD, there may be a cumulative dependence. In this case, the same 1:2 ratio is still expected due to the randomness of $\mathrm{U}$ along with the balanced beamsplitting in path $b$.

Therefore, back to (\ref{eq:cond1/2}), Eve’s detector control attack can be effectively thwarted if Bob uses detectors $D_{ai}$ and $D_{bj}$ for which 
\begin{equation}
\frac{E_{never}^{bj}(I/2)}{E_{never}^{ai}(I)} > \frac{1}{2},
\label{eq:Secure_detect}
\end{equation}
for any value of $I$. We show next that this requirement for Bob’s detectors is realizable in practice.
\\

{
\subsection{Experimental verification of the criteria}
}
We demonstrate that meeting conditions (A) and (B) concurrently can be made impossible in practice by the right choice of Bob’s detectors. 
In our demonstration, we consider an arrangement of two detectors used in the commercial QKD system Clavis2 from ID Quantique, with the values of the threshold parameters obtained from reported results of an experiment by Huang et al. \cite{Huang16}.

Eve’s source [Fig. \ref{fig_setup}(c)] consists of a pulsed laser (vertically polarized, attenuated to $\sim0.6$ pJ/pulse) along with the polarization purity control, unitary polarization transformation, and phase encoding (see Appendix \ref{APP:A:EVE}). The source prepares triggering multi-photon pulses with a time-bin state encoded by Eve’s measured phase $\phi_E$ and a polarization state that can be tuned to any pure or mixed state. The produced state writes:
\begin{equation}
\begin{split}
&(\cos^2 \theta_1 |e \rangle \langle e|+ \sin^2 \theta_1 | \bar{e} \rangle \langle \bar{e}| ) \\
\otimes~ \tfrac{1}{2}  & (| t_l  \rangle + e^{i\phi_E}  |t_s \rangle) (\langle t_l | + e^{-i\phi_E} \langle t_s|).
\end{split}
\label{eq:Eve1}
\end{equation}
with the polarization part be an incoherent mixture of the two arbitrary orthogonal states $|e\rangle $ and $| \bar{e} \rangle$ along with a pure-state time-bin part. 

To assess the alert possibility, the phase $\phi_E$ was set to zero which corresponds to Bob’s detection in the basis state $|D\rangle$ in either path $a$ or $b$. Therefore, the alert possibility due to Eve’s faked-state photons can be analyzed by placing the detectors: $D_{a1}$ and $D_{b1}$ in the alert and secure paths.

Since Eve’s source is able to scan over all points of the Poincaré sphere, there is no loss of generality in fixing the randomizer $\mathrm{U}$ of Bob’s system to a value, unknown to Eve, which we took to be {
\begin{equation}
\mathrm{U}=\frac{1}{\sqrt{2}}
\begin{bmatrix}
i & i \\
1 & -1
\end{bmatrix}.
\end{equation}
This matrix is equivalent to the product of Jones matrices of QWP and HWP, fixed at angles $45^\circ$ and $112.5^\circ$ w.r.t. the vertical axis, respectively. }We used Eve’s source to prepare triggering multiphoton states with purity levels: $\mathcal{P}_T=\{1, 0.78, 0.63, 0.53, 0.5\}$. For each purity setting, HWP2 was rotated from $0^\circ$ to $180^\circ$, with the QWP fixed at $-45^\circ$. During the polarization sweep, the received energies of trigger pulses $E_{a1}$and $E_{b1}$ that reach detectors $D_{a1}$ and $D_{b1}$, respectively, were measured within the superposition time-bin window. 

The threshold function $E_{never} (I)$ is a monotonic increasing function of the blinding power $I$ with a slightly compressive behavior \cite{Huang16}. The requirement in (\ref{eq:Secure_detect}) can be satisfied based on this compressive behavior, and by assigning the detectors of higher sensitivity in the linear mode to the alert path [this higher sensitivity is exhibited by the relatively lower profile of $E_{never} (I)$]. Therefore, based on the measurements of their thresholds (see Appendix \ref{APP:B:TR}), we choose to assign the SPDs $D0$ and $D1$ of Clavis2 system, respectively, to the secure detector $D_{b1}$ and the alert detector $D_{a1}$. 

\begin{figure} [t]
   \begin{center}
   \includegraphics[width=0.7 \linewidth]{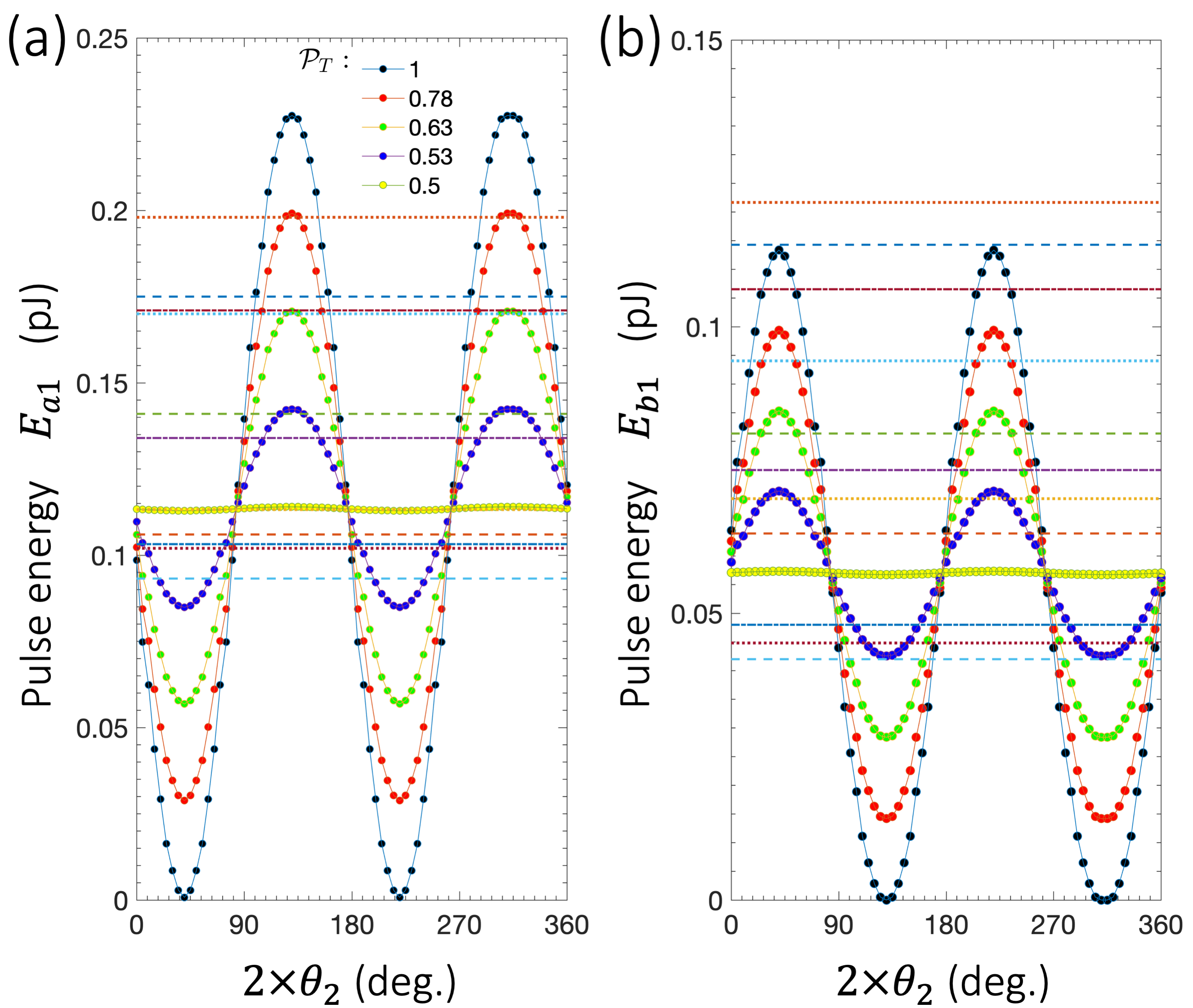}
   \end{center}
   \caption[example] 
   { \label{fig_visibility} 
Measured energies of the triggering pulses: \textbf{(a)} $E_{a1}$ at alert detector $D_{a1}$, and \textbf{(b)} $E_{b1}$at secure detector $D_{b1}$; for five purity levels of Eve's polarization state. The dashed and dotted levels are the detectors thresholds: \textbf{(a)} $E_{never}^{a1} (I_B/4)$, and \textbf{(b)} $E_{never}^{b1} (I_B/8)$ at blinding powers $I_B=\{0.72,0.78,$ $0.86,1.02,1.09,$ $1.27, 1.51,1.78,$ $2.02, 2.26,$ $ 2.5\}$ mW, in bottom-up order. The measurements of pulse energies were selectively performed at the superposition time window. In the measurements, the state is controlled by rotating HWP2 from $0^\circ$ to $180^\circ$ with the QWP fixed at $-45^\circ$. The measured energies was fit to sinusoidal functions of variable visibility. The measurements error is smaller than the marker size.}
 \end{figure}

 Because conditions (A) and (B) rely on the minimum blinding power over all possibilities of $\mathrm{U}$ regardless of the polarization state $\rho_B$, we considered an unpolarized blinding light, without loss of generality. The levels in Fig. \ref{fig_visibility}(a)  and Fig. \ref{fig_visibility}(b) are the thresholds: $E^{a1}_{never} (I_B/4)$ and $E^{b1}_{never} (I_B/8)$, respectively, taken at blinding powers: $I_B=\{0.72,0.78,0.86,1.02,1.09,1.27,1.51,1.78,2.02,2.26,2.5\}$ mW, with the detector gate applied. Eve’s objective is then to find out the blinding power for which the threshold in Fig. \ref{fig_visibility}(a) is greater than the maximum pulse energy received by $D_{a1}$, and concurrently, the corresponding threshold in Fig. \ref{fig_visibility}(b) is less than the maximum pulse energy received by $D_{b1}$ (for the same purity level). 

Figure \ref{fig_visibility} shows the results. It is evident from Fig. \ref{fig_visibility} that Eve cannot meet her objective for any of these levels. Although this is not a complete polarization sweep test (i.e., not covering the entire volume of the Poincaré sphere), it is sufficient to evaluate the ability of a traceless attack. This is because it spans the entire visibility range for arbitrary (pure or mixed) polarization state. Taking into account that $E_{never} (I)$ is a monotonic increasing function of $I$, it can also be verified that this cannot be possible for any other level of blinding power.

\begin{figure} [t!]
   \begin{center}
   \includegraphics[width=9cm]{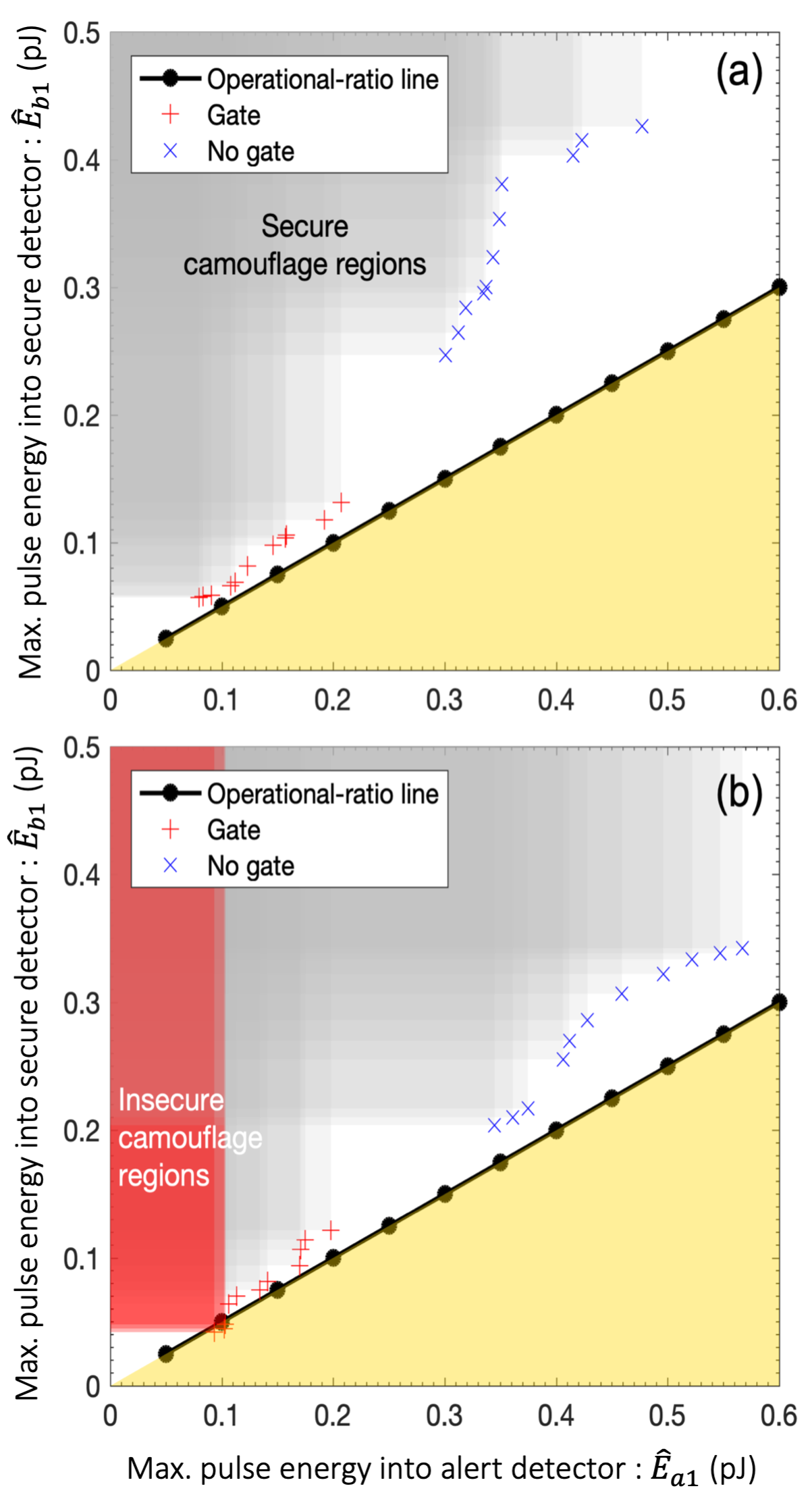}
   \end{center}
   \caption[example] 
   { \label{fig_camouflage} 
Measurements of the maximum pulse energies reaching a secure-path detector: $\hat{E}_{b1}$ and an alert-path detector: $\hat{E}_{a1}$ showing the 1:2 ratio operational line below which the protection fails (yellow area). Intersection points of the thresholds $E_{never}^{a1} (I_B/4)$ (vertical) and $E_{never}^{b1} (I_B/8)$ (horizontal) are shown in the presence (\textcolor{red}{+}) and the absence (\textcolor{blue}{×}) of the detector gate. Each threshold point has a camouflage region (red- and grey-colored areas) within which the maximum pulse energy delivered to $D_{a1}$ is less than $E_{never}^{a1} (I_B/4)$, while the maximum pulse energy for $D_{b1}$ is higher than $E_{never}^{b1} (I_B/8)$. \textbf{(a)} SPD $D1$, which is the more sensitive of the two Clavis2 detectors, is assigned to $D_{a1}$, while the less-sensitive $D0$ is assigned to $D_{b1}$ (see Appendix \ref{APP:B:TR}). In this case all threshold points lie above the operational-ratio line, and the camouflage region does not overlap the unsafe area (marked in yellow). This renders Eve's unnoticeable attack impossible.
 \textbf{(b)} SPDs $D1$ and $D0$ are assigned unwisely to $D_{b1}$ and $D_{a1}$, respectively.  In this case, a small overlap exists between the camouflage area and the unsafe area indicating a possibility for Eve to adjust the parameters: $E_T,\rho_T,I_B,$ and $\rho_B$ and launch a successful attack.   
 }
 \end{figure}

Figure \ref{fig_visibility} shows the results for Eve’s attack using pulses of fixed energy that reach Bob’s system while the detectors gate is applied. Eve may also change the energy level of triggering pulses or launch her attack when the gate of the detector is not applied. Figure \ref{fig_camouflage} shows the results in the presence and absence of the detector gate for a span of trigger pulse energies. It depicts the operational-ratio line which specifies the strict 1:2 relation between the maximum pulse energies reachable to path-$b$ and path-$a$ detectors, as constrained by Bob’s system. 

Figure \ref{fig_camouflage} shows also intersection points between the threshold lines $E_{never}^{a1} (I_B/4)$ and $E_{never}^{b1} (I_B/8)$ (parallel to $y$ and $x$ axes, respectively), combining the thresholds of the two detectors at different values of total blinding power $I_B$. Every intersection point is associated with a camouflage region, where Eve’s detector–control can be enacted tracelessly. As shown in Fig. \ref{fig_camouflage}(a) for a good arrangement of alert and secure detectors, all threshold points are above the operational-ratio line.
This prohibits any overlap between the operational-ratio line of camouflage regions and therefore disallows unnoticeable intrusion. In this arrangement, the necessary and sufficient condition for successful intrusion in (\ref{eq:cond1/2}) is not satisfied for any threshold point. It is then impossible to avoid triggering the alert detectors, no matter what faked-state of light Eve uses.

To show how the unwise choice of Bob’s alert and secure detectors may allow for unnoticeable intrusion, we considered interchanging $D0$ and $D1$ of Clavis2 system to be the alert detector $D_{a1}$ and the secure detector $D_{b1}$, respectively. In this case, some threshold points lied under the operational-ratio line [Fig. \ref{fig_camouflage} (b)]. This creates a valid camouflage region (in overlap with the operational-ratio line) for Eve who can then, in principle, selectively trigger path-$b$ detectors, but not path-$a$ detectors (see Appendix \ref{APP:B:TR}). 



\section{Attack model and security analysis} 
We assume that Eve can introduce photons into Bob’s receiver only through the polarization randomizer.  She is acquainted with the configuration of the system, including timing and other classical information, but has no information on the specific random transformation $\mathrm{U}$ applied at any time. We consider a large number of quantum signals between Alice and Bob, so that all finite-size corrections required in security analysis are negligible (see, e.g., Ref. \cite{Tomamichel12}).

{
Eve interacts identically and independently with each quantum signal. She measures the pulse encoded by Alice in one of the two bases. The outcome of Eve’s measurement is described by three probabilities:  1) $P_e^c \approx \tfrac{1}{2} e^{-\mu(1-F_e)\eta_e} (1-e^{-\mu F_e \eta_e })$ is the probability that Eve’s measurement is in a compatible basis and gives results in a single click in the correct detector. 2) $P_e^w \approx \tfrac{1}{2} e^{- \mu F_e \eta_e} (1-e^{-\mu(1-F_e) \eta_e })$ is the corresponding probability of a click  in the wrong detector only. 3) $P_e^{nc} \approx \tfrac{1}{2} e^{-\tfrac{\mu \eta_e}{2}} (1-e^{-\tfrac{\mu \eta_e}{2}})$ is the probability that Eve’s measurement is in incompatible basis and gives a click in a single detector.  In these expressions, $\mu$ is the mean number of photons per pulse, $F_e$ is the fidelity of Eve’s measurement, and $\eta_e$ is the overall detection efficiency. We specify in the following some possible Eve’s attacks.
}
{
\subsection{Quantum attack}
In this attack, Eve always forwards single-photon pulses to Bob. Bob performs a squashing operation whenever multiple clicks occur \cite{Beaudry08, Tsurumaru08, Gittsovich14}; that is double clicks in different bases do not count, while if in the same basis, they give a random value \cite{Sajeed15}.
}

{
In order to determine the sifted key rate and the QBER under quantum attack, we begin by writing expressions for the raw probabilities $p_{bj} (k)$ that Bob’s detector $D_{bj}, j=1,2,3,4,$ clicks if Eve uses the phase  $k \in \{0,\pi,\pi/2,3\pi/2\}$ to encode her pulse.  For $k=0$,
\begin{equation}
\begin{split}
& p_{b1} (0) \approx c_{b1}+1-\exp{\left(-\frac{\mu_e p_b F \eta_{b1}}{4}\right)},\\
& p_{b2} (0) \approx c_{b2}+1-\exp⁡{\left (-\frac{\mu_e p_b (1-F) \eta_{b2}}{4}\right)},\\
& p_{b3(b4)} (0) \approx c_{b3(b4)} +1- \exp⁡{\left(- \frac{\mu_e  p_b  \eta_{b3(b4)} }{8} \right)},
\end{split}
\label{Eqraw}
\end{equation}
where $c_{bj}$ is the total background rate of detector $D_{bj}$ within the gate slot, $F$ is the fidelity of Bob’s measurement, $\eta_{bj}$ is the overall detection efficiency of $D_{bj}$, and $p_b$ is the probability that Eve’s photon passes into the secure path as given in (\ref{Eqwhichpath}).  Similar expressions apply for other phases $k\in\{\pi,\pi/2,3\pi/2\}$.  
}

{
After the squashing operation, the probability that Bob registers a click in the $D/A$ basis, given that Eve sent a phase-encoded state for $k=0$ is
\begin{equation}
\begin{split}
P_{DA} (0)= &[p_{b1} (0) + p_{b2} (0) - p_{b1}(0) p_{b2}(0)]\\
&\times [1- p_{b3}(0)]  [1- p_{b4}(0)].
\end{split}
\end{equation}
Also, after squashing, the probability that Bob registers a click on $D_{bj}$ given that Eve sent a state coded by the phase $k$ is $P_{bj} (k)$, where, for example,
\begin{equation}
P_{b1} (\pi)=p_{b1} (\pi)[1-\tfrac{1}{2} p_{b2} (\pi)][1-p_{b3} (\pi)]  [1-p_{b4} (\pi)],
\end{equation}
and where $P_{DA} (k)=P_{b1} (k)+P_{b2} (k)$.
}

{
Therefore, given that Alice’s phase $k=0$, the sifted key rate (in path $b$) is
\begin{equation}
\begin{split}
R_b (0) \approx & P_e^c  P_{DA}(0) + P_e^w P_{DA} (\pi)\\
&+ P_e^{nc} [P_{DA}(\pi/2)+P_{DA} (3\pi/2)]\\
& +(1-P_e^c-P_e^w-2P_e^{nc} )  (c_{b1}+c_{b2}-c_{b1} c_{b2})  .
\end{split}
\end{equation}
The corresponding error in Bob’s measurement (in path b) is 
\begin{equation}
\begin{split}
E_b (0) \approx &P_e^c  P_{b2}(0)  +  P_e^w P_{b2} (\pi) \\
&+ P_e^{nc}  [P_{b2} (\pi/2)+P_{b2} (3\pi/2)]\\
&+(1-P_e^c-P_e^w-2P_e^{nc} )  [c_{b2}-(c_{b1} c_{b2})/2].
\end{split}
\end{equation}
}

{
The sifted key rates and the errors – conditioned on Alice’s state with a phase $k\in\{\pi,\pi/2,3\pi/2\}$ – can be similarly obtained.  Consequently, the total sifted key rate and the QBER under Eve’s quantum attack are
\begin{equation}
\begin{split}
&R_b^Q= \frac{1}{4} \sum_{k=\{0,\pi, \frac{\pi}{2} ,\frac{3\pi}{2}\}} R_b (k),\\
&\mathrm{QBER}_b^Q=\frac{1}{4 R_b^Q} \sum_{k=\{0,\pi, \frac{\pi}{2} ,\frac{3\pi}{2}\}} E_b (k).
\end{split}
\end{equation}
Bob does not apply the squashing operation on the alert detections, so that the overall alert rate is
\begin{equation}
\begin{split}
 R_a^Q \approx  c_{a1}+&c_{a2}+2 \\
 - \frac{1}{2} \Big[ &\exp⁡{\left(-\frac{\mu_e p_a F \eta_{a1}}{2}\right)} 
 + \exp⁡{\left(-\frac{\mu_e p_a F \eta_{a2}}{2}\right)} \\
 &+ \exp⁡{\left(-\frac{\mu_e p_a \eta_{a1}}{4}\right)} 
 + \exp⁡{\left(-\frac{\mu_e p_a \eta_{a2}}{4}\right)}  \Big].
\end{split}
\end{equation}
While no obvious change appears in the sifted key rate and the QBER compared to the BB84 protocol, the presence of an alert rate, which is significantly higher than the background rate, provides an additional clear sign of Eve’s attack.
}

{
\subsection{Blinding attack}
In this attack, Eve blinds Bob’s detectors using nonpolarized light, then sends a bright pulse encoded by her measurement outcome. The bright trigger pulse has a pure polarization state. The assumption of nonpolarized blinding light is logical since it is optimal for Eve to render the blinding of all SPDs unaffected by the randomization $\mathrm{U}$. We assume that Eve has complete control over Bob’s measurement in the secure path so that she can limit the QBER; however, as will be shown later, this is not sufficient to limit the rate of the alert. The following analysis is presented in three cases: i) no randomization, ii) randomization, iii) randomization and switching. 
}

{
\textit{i) No randomization.} For simplicity, let us first consider the case: $\mathrm{U}=I$. Aiming to trigger Bob’s secure SPDs in the matched basis and avoid clicks in the unmatched one, Eve sends a trigger pulse energy $E_T$ such that
\begin{equation}
2 E_{never}^{b_j} > \frac{1}{4} E_T > E_{never}^{b_j},     ~~\forall ~j.  
\label{Eqboundj}
\end{equation}
 Note that half the pulse energy $E_T$ will pass while the gate is off (which we assume to result in no action). The lower bound in (\ref{Eqboundj}) assigns a threshold to enable the triggering of matched-basis detectors.  The upper bound puts a limit for not triggering the ones in unmatched basis.
Because an alert detector receives double the blinding power of a secure detector, and due to the compressive nature of $E_{never} (I)$ and the higher sensitivity of alert detectors, we can infer that $E_{never}^{ai} < 2 E_{never}^{bj}, ~\forall ~i,j,$ which can be substituted into the lower bound of (\ref{Eqboundj}) to give 
\begin{equation}
\frac{1}{2} E_T >  E_{never}^{ai},~~\forall ~i.  
\label{Eqboundi}
\end{equation}
The lower bound in (\ref{Eqboundj}) and (\ref{Eqboundi}) signifies that the minimal trigger energy that enables the detectors control in the secure path will also enable triggering of the alert detectors.
}

 \begin{figure} [t!]
   \begin{center}
   \includegraphics[width=10cm]{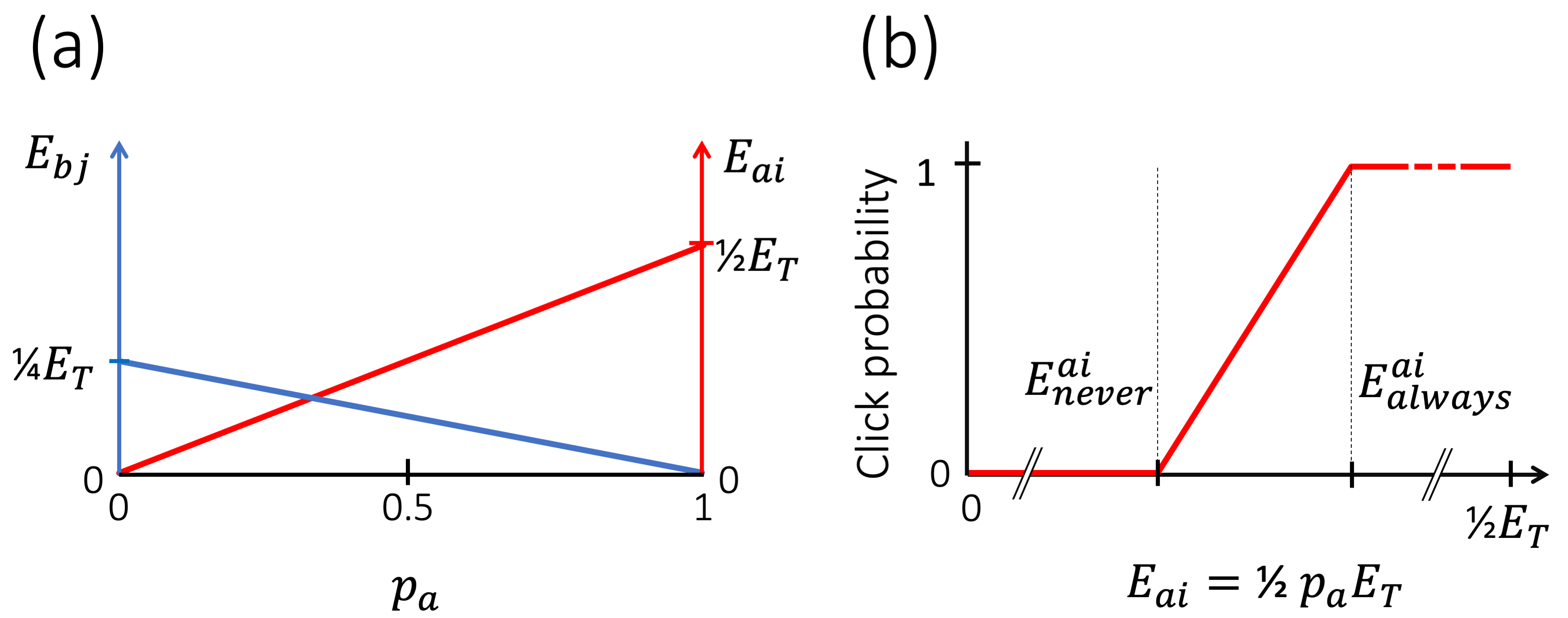}
   \end{center}
   \caption[example] 
   { \label{fig_clickprobability} 
{\textbf{(a)} Portions of the trigger pulse energy $E_T$ that strike two detectors $D_{ai}$ and $D_{bj}$ in matched basis plotted versus the probability $p_a$. \textbf{(b)} Ramp-step approximation of the APD click probability versus the trigger pulse energy under blinding attack. Shown is the case of alert detector $D_{ai}$ when the bases are matched.}
 }
 \end{figure} 

{
\textit{ii) Randomization.} Let us now move to the general case with a random transformation U, but without switching the paths $a$,$b$. In this case, the values of $p_a$ are uniformly distributed between 0 and 1. Figure (\ref{fig_clickprobability})a sketches the energies delivered to SPDs $D_{ai}$ and $D_{bj}$ in matched basis, which equal $\tfrac{1}{2} p_a E_T$ and $\tfrac{1}{4} (1-p_a ) E_T$, respectively.
}

{
For simplicity, we approximate the click probability of blinded detectors by a ramp-step function as plotted in fig. (\ref{fig_clickprobability})b. Therefore, the alert rate on a detector $D_{ai}$ can be obtained by averaging the click probability in fig. (\ref{fig_clickprobability})b over $p_a$ as
\begin{equation}
R_{ai}^{Bl} \approx \frac{1}{2}  \max⁡ \left\{1- \frac{E_{never}^{ai} + E_{always}^{ai}}{E_T}   ,0 \right\},
\label{ratei}
\end{equation}
where the factor $\tfrac{1}{2}$ is attributed to the probability that Eve’s and Bob’s alert-path bases match. 
The trigger rate of a secure detector $D_{bj}$ is obtained similarly by averaging its click probability over $p_a$ as
\begin{equation}
R_{bj}^{Bl} \approx \mathrm{max}⁡ \left\{1- \frac{2(E_{never}^{bj} + E_{always}^{bj})}{E_T}   ,0 \right\},
\label{ratej}
\end{equation}
Therefore, in the absence of switching paths $a$,$b$, the total alert and secure detection rates are 
\begin{equation}
R_a^{Bl}= \frac{1}{2} \sum_{i=1}^2 R_{ai}^{Bl} ,   ~~    R_{secure}^{Bl}=\frac{1}{4} \sum_{j=1}^4 R_{bj}^{Bl}.
\label{noswitchRates}
\end{equation}
}

{
\textit{iii) Randomization and switching.} When Bob switches the alert and secure paths, secure-path detections are counted as alert events and vice versa. If $R_{sw}$ is the switching rate, then the total alert and secure detection rates become
\begin{equation}
\begin{split}
&R_a^{Bl}= \frac{1}{2}(1-R_{sw}) \sum_{i=1}^2 R_{ai}^{Bl} + \frac{1}{4} R_{sw} \sum_{j=1}^4 R_{bj}^{Bl} ,\\
& R_{secure}^{Bl}= \frac{1}{2} R_{sw} \sum_{i=1}^2 R_{ai}^{Bl} + \frac{1}{4}  (1-R_{sw}) \sum_{j=1}^4 R_{bj}^{Bl}.
\end{split}
\label{switchRates}
\end{equation}
This yields the sifted key rate and QBER:
\begin{equation}
R_b^{Bl}=(P_e^c+P_e^w )  R_{secure}^{Bl},
~~\mathrm{QBER}_b^{Bl}= \frac{P_e^w}{P_e^c+P_e^w}.
\label{eqSiftedQBERBlind}
\end{equation}
Remarkably, while this complete blinding attack can keep the level of QBER unaffected by Eve’s interception [as shown by (\ref{eqSiftedQBERBlind})], it is not capable of diminishing the alert rate. Recalling that $E_{never}^{ai} < 2 E_{never}^{b_j} , \forall ~i,j,$ Eqs. (\ref{ratei}), (\ref{ratej}), (\ref{noswitchRates}) show that the alert and secure rates are related by  $R_a^{Bl} \ge\frac{1}{2} R_{secure}^{Bl}$. The alert rate increases proportionally with $R_{sw}$. For example, if $R_{sw}=\frac{1}{2}$, it leads to $R_a^{Bl}=R_{secure}^{Bl}$ as given in (\ref{switchRates}).
}

{
\subsection{Wavelength-dependent blinding attack}
While narrow-band filters can be used to limit a wavelength-dependent attack, it is still possible that Eve elevates the power values of her out-band signals to allow passage of a finite power level \cite{Li2011}.  Such attack may target the (polarizing and non-polarizing) beam splitters and the polarization randomizer. In the former, Eve may exploit the wavelength-dependent deviation from the coupling ratio of the 3-dB coupler and the extinction ratio of the PBS. In the latter, she may exploit the dispersive nature of polarization transformers/controllers, which typically use cascaded birefringent components.  
}

{
We conservatively assume that Eve can develop a wavelength-dependent blinding attack that enables the right control of secure SPDs and always avoids ticking alert SPDs (or at least keeping it below the background rate $c_{ai}$ within the gate slot). Under these conditions, the alert and sifted key rates and QBER are
\begin{equation}
\begin{split}
& R_a^{W|Bl} = R_{sw}, \\
& R_b^{W|Bl} = (P_e^c+P_e^w) (1-R_{sw}),\\
& \mathrm{QBER}_b^{W|Bl}= \frac{P_e^w}{P_e^c+P_e^w}.
\end{split}
\label{EqWBl}
\end{equation}
If Bob keeps $R_{sw}\gg c_{ai},\forall ~ i,$ Eve’s presence will be unveiled by the alert rate. 
}

{
The rates in (\ref{EqWBl}) are also valid for other attack approaches that enable biasing the triggers to the secure detectors. Examples are the attacks exploiting the detector’s efficiency mismatch (e.g., time-shift attacks \cite{Makarov05, Qi07}) or dead time (e.g., the dead-time attack \cite{Henning_Weier11}). Other examples are the pulsed blinding attacks (see, e.g., Refs. \cite{Wu20, Wu-comment20, Henning_Weier11}), where the linear-mode operation of the double-blinded alert detectors last for a longer period \cite{Wu20}; enabling to bias the triggers to secure detectors.
}

{
\subsection{Integrated attacks }
Eve might select one of her menu of attacks at random. If she launches a quantum attack with probability $p_Q$, a blinding attack with probability $p_{Bl}$, or a wavelength-dependent blinding attack with probability $p_{W|Bl}$, then the overall sifted key rate, the QBER, and the alert rate are:
\begin{equation}
\begin{split}
&R_b^e= p_Q  R_b^Q+ p_{Bl} R_b^{Bl}+p_{W|Bl}  R_b^{W|Bl},\\
&\mathrm{QBER}_e=p_Q  \mathrm{QBER}_b^Q + (p_{Bl}+p_{W|Bl} )   \frac{P_e^w}{P_e^c+P_e^w}, \\  
& R_a^e = p_Q  R_a^Q+p_{Bl} R_a^{Bl}+p_{W|Bl} R_a^{W|Bl}.
\end{split}
\end{equation}
}


\section{Discussion and Conclusion}

We have introduced a QKD scheme that nullifies the class of practical hacking strategies exploiting faked-state light, including the detector-control attacks and more generally the intercept-resend strategies. The scheme uses a roundtrip arrangement exploiting the three optical degrees of freedom: polarization, time-bin, and path. Thanks to continuous randomization of the polarization state at the gateway to Bob’s transceiver, only the genuine photon – originally created by Bob – can reliably avoid triggering the alert detectors. We have analytically proven and experimentally verified that this feature can be made unrealizable by Eve’s faked-state light.

It is essential to emphasize that the randomization of the ancillary (polarization) qubit is not by itself sufficient to securely exchange a key without relying on the BB84 protocol to encode the key (time-bin) qubit.  Without BB84, Eve could, in principle, extract the information in the  time-bin qubit in a reliable manner without disturbing the single-photon state forwarded to Bob.

\section{Appendices} 
\subsection{Eve’s state preparation}\label{APP:A:EVE}
 

 In order to generate light pulses with a prepared state of polarization, mimicking that potentially employed by Eve, we have used the three-stage optical system in Fig. \ref{fig_setup}(c). The first stage produces mixed-state pulses with a polarization purity set by a half-wave plate HWP1 followed by a heavily unbalanced polarization-based Mach-Zehnder interferometer (PMZI). The propagation times through the two PMZI arms differ by the period of the pulsed laser, which is longer than its coherence time. The PMZI thus mixes pairs of mutually incoherent pulses of orthogonal polarization with a ratio set by the rotation angle $\theta_1$ of HWP1. 
 The polarization purity is then given by 
 \begin{equation}
 \mathcal{P}_T=1 -  \tfrac{1}{2} \sin^2  4\theta_1 ,
 \end{equation} with the values $\mathcal{P}_T=\{1, 0.78, 0.63, 0.53, 0.5\}$ used in the experiment corresponding to HWP1 angles: $\theta_1=\{0^\circ, 10.4^\circ, 15^\circ, 18.9^\circ, 22.5^\circ\}$. 
 
\begin{figure} [t]
 \begin{center}
 \includegraphics[width=10cm]{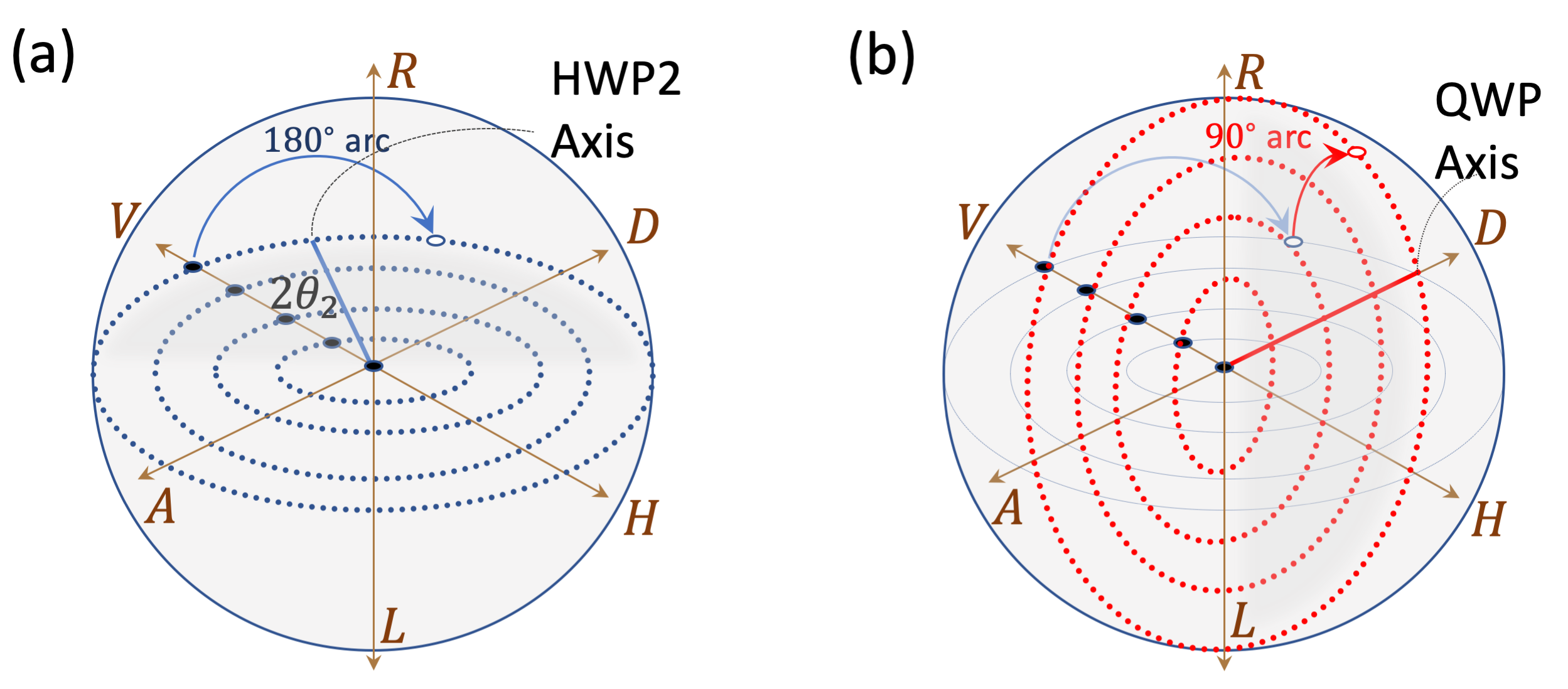}
  \end{center}
  \caption[example] 
   { \label{fig_Poincare} 
{Evolution of the state of polarization (SOP) of Eve’s faked state on the Poincaré sphere. \textbf{(a)} Mixed state is generated with different purities by rotating HWP1 (black dots).  Rotation of HWP2 from $0^\circ$ to $180^\circ$ moves the SOP to span the blue dotted circles in the plane orthogonal to the $|R\rangle$-$|L\rangle$ axis \textbf{(b)} QWP with axis at $-45^\circ$ performs $90^\circ$rotation in the plane orthogonal to the $|D\rangle$-$|A\rangle$ axis (red dotted circles). Arbitrary transformation can be done by rotating HWP2 and QWP. Eve's transformation arrangement along with the measurement in Bob's system work in a way similar to a de Sénarmont compensator. }}
 \end{figure}

The second stage of the system uses a half-wave plate HWP2 (its rotation angle is $\theta_2$) and a quarter-wave plate QWP to perform the unitary rotation over the Poincaré sphere (see Fig. \ref{fig_Poincare}) and create the polarization state in (\ref{eq:Eve1}). After preparing the polarization state, the third stage creates phase-encoded time bin state using a Mach-Zehnder interferometer (MZI) identical to the one used by Bob. 

\subsection{Triggering thresholds of “Clavis2” SPDs}\label{APP:B:TR}
\begin{figure} [t]
   \begin{center}
   \includegraphics[width=10cm]{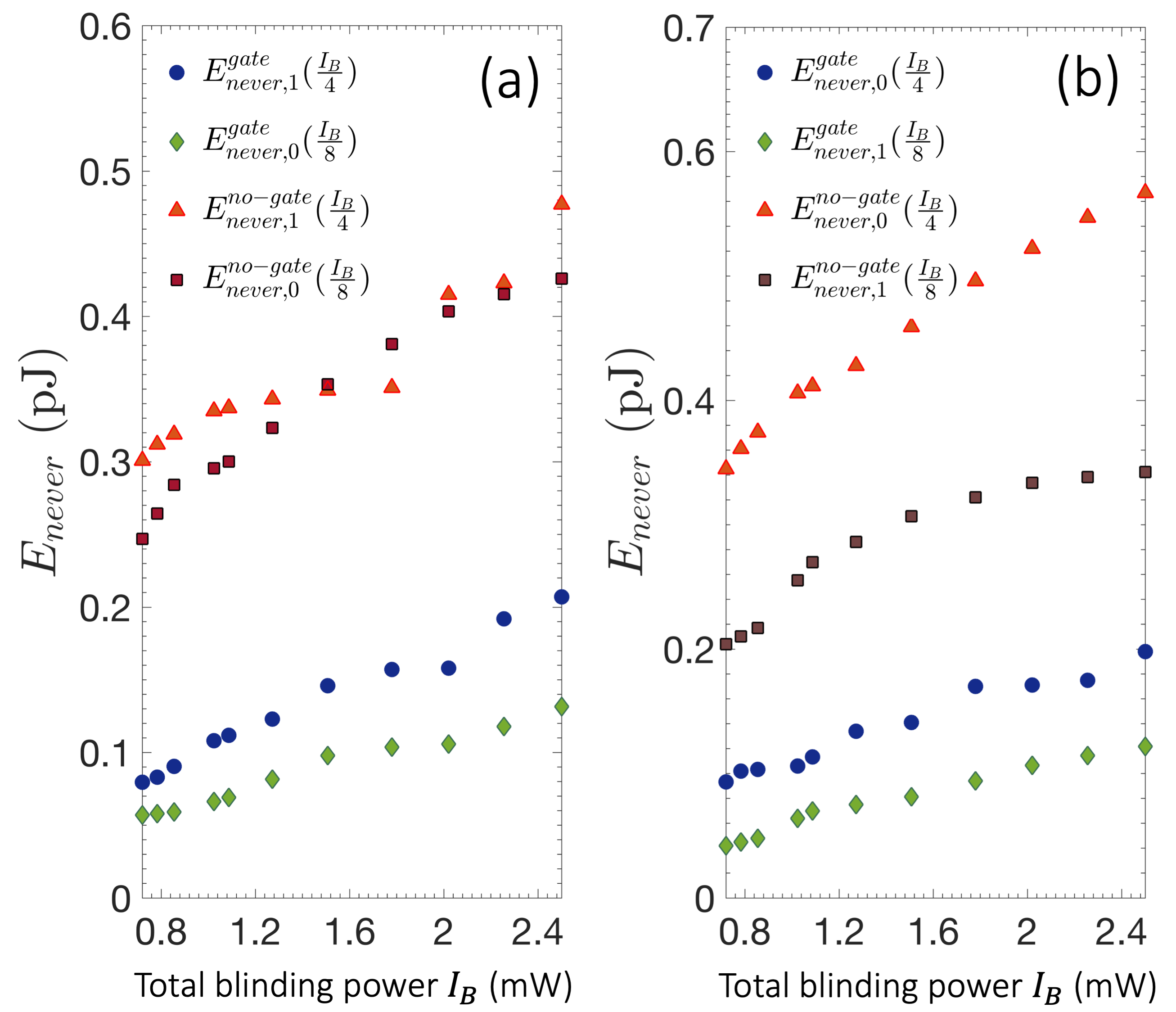}
   \end{center}
   \caption[example] 
   { \label{fig_Thresholds} 
Thresholds $E_{never}$ of the two SPDs $D0$  and $D1$ used in the Clavis2 system in the presence and the absence of the gate, plotted versus the blinding power $I_B$ of unpolarized light injected into Bob’s system. \textbf{(a)} $D0$ and $D1$ are assigned to paths $b$ and $a$, respectively. \textbf{(b)} $D0$ and $D1$ are switched to paths $a$ and $b$.
 }
 \end{figure}

Figures \ref{fig_visibility} and \ref{fig_camouflage} depict the threshold values: $E_{never,0}^{gate}, E_{never,1}^{gate}, E_{never,0}^{no-gate}$, and $E_{never,1}^{no-gate}$ for the two Clavis2 detectors $D0$ and $D1$ at different values of blinding power. This data was reproduced from the experimental results in \cite{Huang16} and supplemented by interpolations to deduce some missed points.  Figure \ref{fig_Thresholds}(a) shows these thresholds versus the power $I_B$ (total Eve’s blinding power input to Bob’s system), when the two SPDs $D1$ and $D0$ are inserted into the alert path $a$ and the secured path $b$, respectively. Figure \ref{fig_Thresholds}(b) shows the other unwise alternative when $D0$ and $D1$ are in path $a$ and path $b$, respectively. It is obvious that the condition in (\ref{eq:cond1/2}), which is necessary and sufficient for a traceless attack, is not satisfied at all points in the first case of Fig. \ref{fig_Thresholds}(a). This verifies the security of Bob’s system against Eve’s detector-side attack.
By contrast for the alternative arrangement, the condition (\ref{eq:cond1/2}) is satisfied at some points (particularly, the first three points of gated detection) of Fig. \ref{fig_Thresholds}(b). This enables the overlap between the camouflage regions of these points and the operational-ratio line and allows for a traceless detector-side attack by Eve, as depicted in Fig. \ref{fig_camouflage}(b). 

\subsection{Bounds for the overlap between pure and mixed states}\label{APP:C:Overlap}
We show here that the maximum and minimum overlaps between a pure state $|\psi\rangle$ and a mixed state $\rho$ after the application of an arbitrary unitary operator $\mathrm{U}$, are given by
\begin{equation}
\begin{split}
\max \{\langle \psi| \mathrm{U} \rho \mathrm{U}^+ |\psi\rangle\}= \frac{1}{2} \left(1+ \sqrt{2\mathcal{P}-1} \right),\\
\min \{\langle \psi| \mathrm{U} \rho \mathrm{U}^+ |\psi\rangle\}= \frac{1}{2} \left(1-\sqrt{2\mathcal{P}-1} \right),
\label{eq:maxmin}
\end{split}
\end{equation}
where $\mathcal{P}$ is the purity of the state $\rho$. {Let us express the mixed state $\rho$ as a mixture $\lambda |v \rangle \langle  v | + (1-\lambda) |\bar{v}\rangle \langle \bar{v}|$ of two orthogonal states $|v\rangle$ and $|\bar{v}\rangle$, then, after applying $\mathrm{U}$, the overlap with the state $|\psi \rangle$ writes
\begin{equation}
\begin{split}
\langle \psi| \mathrm{U} \rho \mathrm{U}^+ | \psi \rangle &= \langle \psi|\mathrm{U} ~\big[\lambda |v \rangle \langle  v | + (1-\lambda) |\bar{v}\rangle \langle \bar{v}|\big]~\mathrm{U}^+|\psi\rangle \\
&= \lambda \left|\langle \psi| \mathrm{U} |v \rangle \right|^2  + (1-\lambda) \left|\langle \psi |\mathrm{U} |\bar{v} \rangle \right|^2.
\end{split}
\label{eq: interpolate}
\end{equation}
Since $ |\langle \psi| \mathrm{U} |v \rangle |^2  =1- |\langle \psi |\mathrm{U} |\bar{v} \rangle |^2$, Eq. (\ref{eq: interpolate}) describes the overlap as an interpolation between the two complementary probabilities $\lambda$ and $(1-\lambda)$, where the interpolation weights $ |\langle \psi| \mathrm{U} |v \rangle |^2$ and $|\langle \psi |\mathrm{U} |\bar{v} \rangle |^2$ vary according to $\mathrm{U}$. Hence the probabilities $\lambda$ and $(1-\lambda)$ give the maximum and minimum overlaps between the states $\mathrm{U} \rho \mathrm{U}^+$ and $|\psi\rangle$. The purity of $\rho$ is $\mathcal{P}=\mathrm{tr}(\rho^2 )= \lambda^2+(1-\lambda)^2$, which offers two values for $\lambda$ based on $\mathcal{P}$, and leads directly to the two bounds in (\ref{eq:maxmin}).}

\end{document}